# Real-Time Glaucoma Detection from Digital Fundus Images using Self-ONNs

Ozer Can Devecioglu[1], Junaid Malik[1], Turker Ince[2], Serkan Kiranyaz[3], *Senior Member, IEEE* Eray Atalay[4], and Moncef Gabbouj[1], *Fellow, IEEE.*

[1] Department of Computing Science, Tampere University, Tampere, Finland
[2] Department of Electrical & Electronics Engineering, Izmir University of Economics, Izmir, Turkey
[3] Electrical Engineering Department, Qatar University, Doha, Qatar
[4] Department of Ophthalmology, Faculty of Medicine, Eskişehir Osmangazi University, Eskişehir

Corresponding author: Ozer Can Devecioglu (e-mail: ozer.devecioglu@tuni.fi).

**ABSTRACT** Glaucoma leads to permanent vision disability by damaging the optical nerve that transmits visual images to the brain. The fact that glaucoma does not show any symptoms as it progresses and cannot be stopped at the later stages, makes it critical to be diagnosed in its early stages. Although various deep learning models have been applied for detecting glaucoma from digital fundus images, due to the scarcity of labeled data, their generalization performance was limited along with high computational complexity and special hardware requirements. In this study, compact Self-Organized Operational Neural Networks (Self-ONNs) are proposed for early detection of glaucoma in fundus images and their performance is compared against the conventional (deep) Convolutional Neural Networks (CNNs) over three benchmark datasets: ACRIMA, RIM-ONE, and ESOGU. The experimental results demonstrate that Self-ONNs not only achieve superior detection performance but can also significantly reduce the computational complexity making it a potentially suitable network model for biomedical datasets especially when the data is scarce.

## I. INTRODUCTION

Glaucoma, also called "the silent thief of sight," leads to permanent vision disability by damaging the optic nerve. According to the World Health Organization (WHO) data, glaucoma is the leading cause of irreversible blindness globally[1]. Because the optic nerve head damage caused by glaucoma is irreversible, early diagnosis and treatment is crucial. However, mild glaucoma does not show any symptoms such as pain or blurred vision, hence its detection can be challenging especially for large-scale screening purposes. Although the previous worldwide estimate of the number of adults with glaucoma was 64.3 million in 2013, projections show that this figure will rise by 74% to 111.8 million in 2040 [2].The optic nerve head damage in glaucoma can be diagnosed using various clinical tools including but not limited to fundoscopy, visual field examination, optical coherence tomography and digital fundus imaging. Recently, thanks to its non-invasive, cost-effective, and rapid nature, digital fundus images have been proposed as an effective means of exploiting signal processing and machine learning techniques for the automated assessment of the optic nerve head in a large-scale glaucoma screening setting, e.g., see Figure 1.

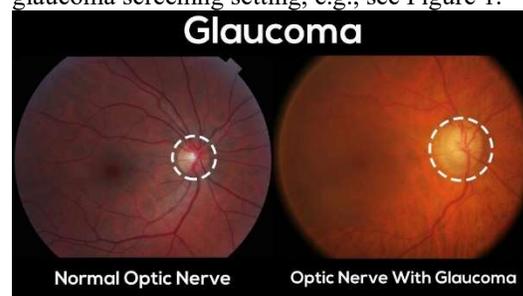

**FIGURE 1:** Optic nerves in glaucoma and normal eye [2].

Several methods for the automatic detection of glaucoma have been proposed in the literature. Bock *et al*. [4] transformed color fundus images to eigenimages by principal component analysis and classified using a support vector machine (SVM) to obtain a Glaucoma Risk Index (GRI) with competitive Glaucoma detection performance. Dua *et al*. [5] suggested a glaucoma detection system using wavelet transform features to extract energy signatures and applied different feature ranking and feature selection strategies. They classified these features by SVM and obtained 93% accuracy using a local dataset. Carillo *et al*. [6] proposed a computational tool based on the optic disc (OD) and cup segmentation algorithm for estimating the cup-to-disc ratio (CDR) and thresholding it for automatic glaucoma detection. They obtained 88.5% classification accuracy over the set of fundus images gathered in the Center of Prevention and Attention of Glaucoma in Bucaramanga. In [7], Nayak *et al*. presented a glaucoma detection method using digital image processing techniques. They applied pre-processing, morphological operations and thresholding for automatically detecting the OD, the blood vessels and computing the features. These features are validated by classifying the normal and glaucoma fundus images collected at the Kasturba Medical College, India using a neural network classifier. Their system achieved a sensitivity of 100% and a specificity of 80% over the test dataset. Singh *et al*. [8] developed an algorithm for extracting blood vessels from a fundus image and applied SVM to classify the wavelet features of the segmented OD image, achieving 94.7% accuracy for glaucoma detection from their local dataset. Recently, the focus has particularly been drawn to deep learning for glaucoma detection [9]. In [10], Ajitha and Judy proposed a glaucoma detection system by applying a faster region-based CNN (R-CNN) which can extract the regions of interest from the fundus images. The pre-trained deep models, ResNet50 and VGG16, are used to classify them. For the DRISHTI dataset, they reported accuracy scores of 92.5% and 92% for these models, respectively. For the ORIGA dataset, they obtained approximately 90% accuracy using the pre-trained models. In [11], Fu et al. evaluated two different deep learning-based glaucoma assistive diagnosis methods: the OD and OC segmentation-based multi-label segmentation network (M-Net) and disc-aware ensemble network (DENet) which can learn to predict glaucoma directly from the fundus images. DENet and M-Net produced 84.29% and 81.57% classification accuracy, respectively with the ORIGA dataset. DENet resulted in a better performance when there is sufficient training data with similar image distribution; otherwise, the segmentation-based M-Net method produced higher performance. In, [12] Ahn *et al*. proposed a method to detect glaucoma using deep CNNs. They trained and tested their system using 1542 fundus images collected at Kim's Eye Hospital, Japan, and obtained overall 87.9% classification accuracy. In [13], Sreng *et al*. proposed an automatic two-stage glaucoma screening system based on two separate ensembles of deep CNNs to first segment the OD and then use as input to classify glaucoma disease from the REFUGE, ACRIMA, ORIGA, RIM-ONE, and DRISTI-GS1 benchmark datasets with the accuracy of 97.37%, 90.00%, 86.84%, and 99.53%, respectively.

There are several drawbacks and limitations of applying a deep CNN model to this problem. First of all, it is known that the performance of deep CNNs suffers from data scarcity which is common in several biomedical applications. To partially remedy this drawback, several studies resort to dropout and data augmentation. Additionally, such deep networks have high computational complexity and require specialized hardware, both of which prevent their usage in low-power computing environments in real-time. Reducing the network complexity would of course be an obvious solution for this; at the expense of a performance drop. More important drawbacks have been identified by recent studies [14]-[21] which pointed out that CNNs are homogenous network models that use a simplistic "linear" neuron model. The latter is known to be an oversimplistic and crude model of the biological neuron. On the other hand, the mammalian neural or visual systems, for instance, are highly heterogeneous and composed of highly diverse neuron types with distinct biochemical and electrophysiological properties [26]-[31]. This is why conventional homogenous networks such as CNNs fail to learn the problems whenever the solution space is highly nonlinear and complex [14]-[21] unless a sufficiently high network depth and complexity (variants of CNN) are accommodated.

To address these limitations, Operational Neural Networks (ONNs)[21]-[24] have recently been proposed as a heterogeneous network model encapsulating distinct non-linear neurons. ONNs are derived from the Generalized Operational Perceptrons (GOPs) [14],[20] that can learn those problems where MLPs entirely fail. Following GOPs footsteps, ONNs have outperformed CNNs significantly and succeeded in some complex problems where CNNs entirely failed. Recently, a new variant of ONNs, called Self-Organized Operational Neural Networks (Self-ONNs[1]), have been proposed. Unlike ONNs, Self-ONNs do not require any prior operator search. Instead, during the training of the network, to maximize the learning performance, each generative neuron in a Self-ONN can customize the nodal operators, $\Psi$, of each kernel connection. This yields a heterogeneity level that is far beyond that of ONNs, and the traditional "weight optimization" becomes an "operator generation" process as the details can be found in [24]. The superior regression capability of Self ONNs over image segmentation, restoration, and denoising was demonstrated in recent studies; however, they have not been evaluated for a classification problem.

In this study, we propose compact Self-ONNs for glaucoma detection in fundus images and evaluate their performance extensively over the three benchmark datasets. This is the first study where Self-ONNs are evaluated against deep CNNs over a classification problem. Moreover, this is the first time a novel and compact network model has ever been proposed and evaluated against deep CNNs on a biomedical dataset. The main motivation behind this is to demonstrate the fact that SelfONNs, as highly heterogeneous networks with the generative neuron model can achieve and even surpass the diversity level of the deep CNNs even with a compact configuration. Especially when the data is scarce, this will not only yield a superior classification performance level, also enables an elegant computational efficiency.

---

[1] The optimized PyTorch implementation of Self-ONNs is publically shared in http://selfonn.net/ .

The rest of the paper is organized as follows: Self-ONNs are briefly reviewed in Section II. The proposed glaucoma detection framework and the experimental results are discussed in Section III, where the performance of the proposed technique is assessed against conventional state-of-the-art CNN-based approaches. Finally, Section IV concludes the paper and suggests topics for future research.

## II. THE PROPOSED APPROACH

### A. Self-organized Operational Neural Networks
In this section, we will briefly summarize Self-ONNs and their main properties. Self-ONNs are formulated using a nodal transformation, $\psi$, based on the Taylor-series function approximation near the origin ($a=0$),

$$\psi(x) = \sum_{n=0}^{\infty} \frac{\psi^{(n)}(0)}{n!} x^n \qquad (1)$$

The $Q^{th}$ order truncated approximation, formally known as the Taylor polynomial, takes the form of the following finite summation:

$$\psi(x)^{(Q)} = \sum_{n=0}^{Q} \frac{\psi^{(n)}(0)}{n!} x^n \qquad (2)$$

The above formulation can approximate any function $\psi(x)$ sufficiently well near 0. When the activation function bounds the neuron's input feature maps in the vicinity of 0 (e.g., *tanh*) the formulation of (2) can be exploited to form a composite nodal operator where the power coefficients, $\frac{\psi^{(n)}(0)}{n!}$ can be the learned parameters of the network during training. It was shown in [23] that the nodal operator of the $k^{th}$ generative neuron in the $l^{th}$ layer can take the following general form:

$$\psi_l^k(Y_{l-1}, W_l^k, Q) = \sum_{q=1}^{Q} Y_{l-1}^q \otimes W_l^{k\,(q)} \qquad (3)$$

where $Y_{l-1}^q$ is the corresponding input and $W_l^k \in \mathbb{R}^{\widehat{M} \times \widehat{N} \times Q}$ is the three-dimensional weight matrix and $W_l^{k(q)} \in \mathbb{R}^{\widehat{M} \times \widehat{N}}$ is the $q^{th}$ slice of $W_l^k$. The $0^{th}$ order term $a$, the DC bias, is ignored as its additive effect can be compensated by the learnable bias parameter of the neuron.

More detailed information about the theory and forward-propagation formulations of Self-ONNs can be found in the [23].

### B. The Proposed Glaucoma Diagnosis Framework
The general framework of the proposed automated glaucoma detection scheme is shown in Figure 2. where Self-ONNs analyze the normalized RGB digital fundus images of size 128x128. Each color channel (RGB) of the input image is resampled to 128x128 and normalized by linear scaling to the range of [-1. 1], as follows:

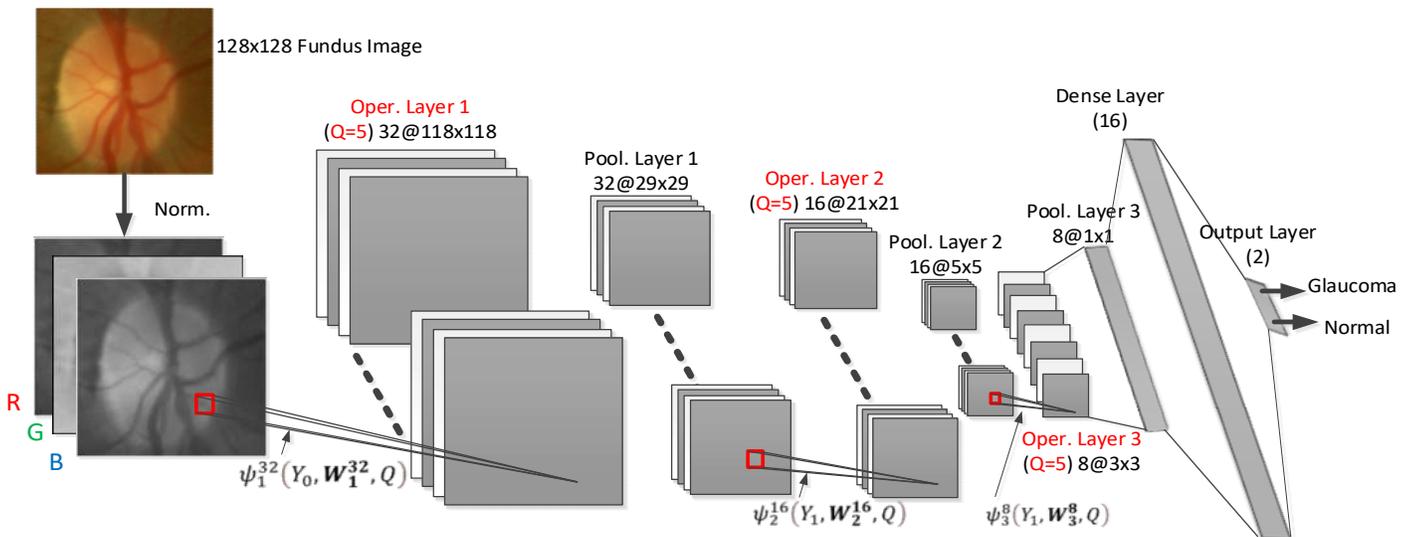

FIGURE 2: The proposed glaucoma diagnosis framework.

$$X_N(i,j) = 2\frac{X(i,j) - X_{min}}{X_{max} - X_{min}} - 1 \qquad (4)$$

where $X(i,j)$ and $X_N(i,j)$ are the original and normalized pixel values, $X_{max}$ and $X_{min}$ are the maximum and minimum values of the input color channel, respectively. As illustrated in the figure, the Self-ONN architecture has 32, 16, and 8 neurons in the three hidden layers, respectively, and through self-organization of its nodal operators it can perform the required non-linear transformations to extract optimal features from the raw fundus images. There are 16 neurons in the dense (MLP) layer for classification. The input layer size is 3 corresponding to the three normalized raw input color channels and the output layer size is 2 corresponding to the number of classes. Through the BP training of Self-ONNs as explained in the Appendix, the optimal non-linear operators can be learned to maximize the learning performance and achieve a superior classification of Glaucoma disease. The nonlinear activation function tanh is used in Self-ONNs. The kernel sizes are set as 11x11, 9x9, and 3x3, respectively. The sub-sampling factors for the pooling layers are set as 4, 4, and 2, respectively. In the figure, Q is set to 5 for all operational layers; however, to investigate the effect of nonlinearity, Self-ONNs with 4 different Q values in the range of [3, 9] are configured and tested. As stated earlier, CNN is a special case of Self-ONN with $Q = 1$ assigned for all neurons. Therefore, the experimental setup shown in Figure 2 can also conveniently be also used for evaluating conventional CNNs. The experimental setup and network parameters will be presented in the next section.

## III. EXPERIMENTAL RESULTS

In this section, we shall first outline the three benchmark glaucoma datasets used in this study and present the experimental setup used for testing and evaluation of the proposed Self-ONNs based glaucoma detection. We shall then present the overall results obtained from the glaucoma detection experiments and perform comparative evaluations against several competing techniques. Additionally, the computational complexity of the proposed method for both training and classification will be evaluated in detail. The three benchmark datasets, ACRIMA, RIM-ONE, and ESOGU are used for training and testing of the proposed

### A. Glaucoma Diagnosis Benchmark Datasets

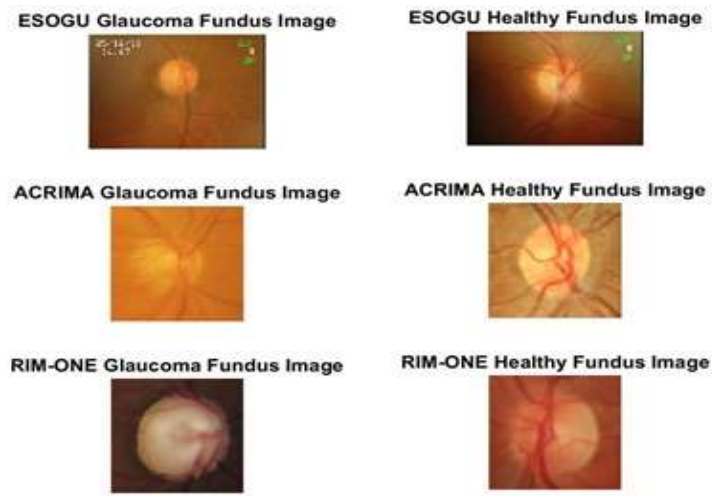

**FIGURE 3:** Sample glaucoma and healthy images from ESOGU, ACRIMA, and RIM-ONE datasets.

In Table I, the details of the three benchmark datasets used to evaluate the performance of the proposed glaucoma assistive diagnosis system are presented. The ESOGU dataset [37] consists of 4725 optic nerve photographs of normal and glaucoma patients at the Ophthalmology Clinic, at the Faculty of Medicine, Eskişehir Osmangazi University in Turkey. This dataset was collected under the Declaration of Helsinki and approved by the Eskişehir Osmangazi University's Ethics Committee. All photographs were acquired by the non-mydriatic fundus camera (Kowa nonmyd alpha-DIII). All 20° posterior segment photographs in the image archive were extracted after clearing the identity and gender information. Qualities of the fundus images and qualities were classified in terms of sharpness. The database is labeled by three expert doctors according to the instructions in [33]-[36]. The commonly used ACRIMA [2] and RIM-ONE [32] datasets are composed of 705 and 455 fundus images, respectively. Sample glaucoma and healthy images from each of the ESOGU, ACRIMA, and RIM-ONE datasets are shown in Figure 3.

**Table I: Benchmark Datasets.**

| Name | Healthy Fundus Image # | Glaucoma Fundus Image # |
|---|---|---|
| ESOGU [37] | 4405 | 320 |
| ACRIMA [2] | 309 | 396 |
| RIM-ONE [32] | 261 | 194 |

### B. EXPERIMENTAL SETUP

In this study, we propose compact Self-ONNs for glaucoma detection and performed an extensive set of evaluations over the three benchmark datasets. As the first study where Self-ONNs are evaluated against deep CNNs over a classification problem, we have performed both fair (with equivalent configurations) and unfair (compact Self-ONNs vs deep CNNs) comparisons. For the latter, we have even used deep CNN models pre-trained and then applied transfer learning on the benchmark datasets. Moreover, to prevent a potential bias or boost from another (3rd party) methodology, we keep the configuration and training of Self-ONNs as "default" as possible, i.e., no dropout, data augmentation, residual blocks, parameter tuning, etc. This is also true for the dense layers that are naturally attached to the Self-ONNs for classification: conventional MLP hidden and output layers are used instead of their heterogeneous alternative such as GOP's operational layers. In this way, the net contribution of Self-ONN's operational layers with the generative neuron model can be assessed. Additionally, we used the compact Self-ONN illustrated in Figure 2 to achieve high computational efficiency for training and particularly for real-time detection.

We perform fair comparisons against CNN models with equivalent configuration as well as unfair ones against some of the state-of-the-art deep networks such as ResNet-101 [38] and VGG-19 [39]. In the latter, we used pre-trained deep CNN models and then applied transfer learning on the benchmark datasets. Moreover, to assess the performance of the proposed Self-ONNs, we avoid using dropout, data augmentation, residual blocks, and parameter tuning. For the dense layers following the Self-ONN layers, conventional MLP hidden and output layers are used instead of their heterogeneous alternative such as GOP layers. In this way, the net contribution of Self-ONNs layers with the generative neuron model can be assessed.

For all experiments, we employ a shallow training scheme with a maximum of 50 BP iterations. The other stopping criterion is the minimum train classification error level, which is set to 3% to prevent over-fitting. We initially set the learning rate, ε, as 10-4 and used the Adam optimizer. The mean-squared error (MSE) is used as the loss function. A 10-fold cross-validation technique is applied over the three benchmark datasets separately to prevent overfitting and better estimate the generalization performance of the classifier. For each fold, 5 BP runs are performed and the best classification performance is reported.

### C. PERFORMANCE EVALUATIONS

The following commonly used performance metrics are used: Accuracy, Balanced Accuracy, Precision or Positive Predictivity (Ppr), Recall, or Sensitivity (Sen), Specificity, F1-score (F1), F2-score (F2). These metrics are distinctive for each class and they assess the capability of the proposed classifier to distinguish specific events from non-events. The formulations for these performance metrics in terms of false negatives (FN), false positives (FP), true negatives (TN), and true positives (TP) can be expressed as follows:

$$Acc = \frac{TP + TN}{TP + FP + TN + FN}$$

$$Balanced\ Acc = \frac{R + Spe}{2},$$

$$R = \frac{TP}{TP + FN}, Spe = \frac{TN}{TN + FP} \quad (5)$$

$$P = \frac{TP}{TP + FP},\quad F1 = \frac{2PR}{P + R}$$

$$F2 = (1 + 2^2)\frac{PR}{2^2 P + R}$$

Table II, Table III and Table IV report the classification results of the proposed Self-ONN-based framework for the ESOGU, ACRIMA, and RIM datasets, respectively. As mentioned earlier, the performance of four different Self-ONN models with Q values set as 3, 5, 7 and 9 are compared against the equivalent CNN models and ResNet-101 and VGG-19 deep networks. The ResNet-101 and VGG-19 networks were pre-trained, and they were fine-tuned over each dataset using transfer learning. All digital fundus images were resized to 128x128, normalized and linearly scaled prior to classification.

There are several important observations worth mentioning over the results. For the ESOGU dataset, Self-ONN with Q=3 achieves the best (100%) F1-score, outperforming an equivalent CNN by 8%. The same Self-ONN also outperforms the deep CNN models, ResNet-101 and VGG. The superior performance of Self-ONN with an even higher gain is observed for the RIM-ONE dataset, where Self-ONN with Q=5 achieved the highest F1-score of 74%, which is 12% higher than CNN with the same configuration. VGG-19 model, which was trained using transfer learning achieves the second-best result with a 73%

TABLE II Glaucoma classification performances on the ESOGU dataset. The best F1-score and F2-score are highlighted in bold.

| Network | Configuration | Accuracy (%) | Balanced Accuracy (%) | Sensitivity (%) | Specificity (%) | Precision (%) | F1-Score (%) | F2-Score (%) |
|---|---|---|---|---|---|---|---|---|
| Eqv. CNN | (32+16+8) + (16+2) | 94.1 | 93.0 | 87.7 | 98.4 | 97.4 | 92.3 | 89.4 |
| Self-ONN (Q=3) | (32+16+8) + (16+2) | 100.0 | 100.0 | 100.0 | 100.0 | 100.0 | **100.0** | **100.0** |
| Self-ONN (Q=5) | (32+16+8) + (16+2) | 94.7 | 94.0 | 90.9 | 97.2 | 95.6 | 93.2 | 91.8 |
| Self-ONN (Q=7) | (32+16+8) + (16+2) | 98.3 | 98.4 | 99.0 | 97.8 | 96.8 | 97.9 | 98.5 |
| Self-ONN (Q=9) | (32+16+8) + (16+2) | 95.0 | 94.5 | 91.8 | 97.2 | 95.7 | 93.7 | 92.5 |
| ResNet-101* | 101-Layer | 97.4 | 97.3 | 98.4 | 96.2 | 97.1 | 97.7 | 98.1 |
| VGG-19* | 19 Layer | 98.9 | 98.7 | 100.0 | 97.5 | 98.1 | 99.0 | 99.6 |

*Transfer learning is used.

TABLE III Glaucoma classification performances on the RIM-ONE dataset. The best F1-score and F2-score are highlighted in bold.

| Network | Configuration | Accuracy (%) | Balanced Accuracy (%) | Sensitivity (%) | Specificity (%) | Precision (%) | F1-Score (%) | F2-Score (%) |
|---|---|---|---|---|---|---|---|---|
| Eqv. CNN | (32+16+8) + (16+2) | 68.0 | 68.3 | 51.3 | 85.45 | 78.6 | 62.1 | 55.1 |
| Self-ONN (Q=3) | (32+16+8) + (16+2) | 71.5 | 71.7 | 62.1 | 81.36 | 77.7 | 69.0 | 64.6 |
| Self-ONN (Q=5) | (32+16+8) + (16+2) | 75.3 | 75.4 | 68.2 | 82.73 | 80.5 | **73.9** | 70.3 |
| Self-ONN (Q=7) | (32+16+8) + (16+2) | 74.0 | 74.1 | 65.6 | 82.73 | 79.8 | 72.0 | 68.0 |
| Self-ONN (Q=9) | (32+16+8) + (16+2) | 73.7 | 73.9 | 66.0 | 81.82 | 79.1 | 72.0 | 68.2 |
| ResNet-101* | 101-Layer | 69.8 | 65.1 | 33.1 | 97.2 | 87.3 | 48.0 | 37.7 |
| VGG-19* | 19 Layer | 77.3 | 76.8 | 71.6 | 82.0 | 73.9 | 72.7 | **72.0** |
| Xception*[2] | 36-Layer | 71.2 | 79.6 | 79.3 | 79.9 | N/A | N/A | N/A |

*Transfer learning is used.

**TABLE IV** Glaucoma classification performances on the ACRIMA dataset. The best F1-score and F2-score are highlighted in bold.

| Network | Configuration | Accuracy (%) | Balanced Accuracy (%) | Sensitivity (%) | Specificity (%) | Precision (%) | F1-Score (%) | F2-Score (%) |
|---|---|---|---|---|---|---|---|---|
| *Eqv. CNN* | (32+16+8) + (16+2) | 80.8 | 79.3 | 85.2 | 73.4 | 84.4 | 84.8 | 85.0 |
| *Self-ONN (Q=3)* | (32+16+8) + (16+2) | 91.8 | 91.2 | 93.6 | 88.8 | 93.4 | 93.5 | 93.5 |
| *Self-ONN (Q=5)* | (32+16+8) + (16+2) | 94.5 | 93.4 | 94.5 | 92.4 | 93.3 | **93.9** | **94.3** |
| *Self-ONN (Q=7)* | (32+16+8) + (16+2) | 86.1 | 84.5 | 90.6 | 78.4 | 87.6 | 89.1 | 90.0 |
| *Self-ONN (Q=9)* | (32+16+8) + (16+2) | 81.4 | 77.7 | 92.05 | 63.4 | 81.0 | 86.1 | 89.6 |
| *ResNet-101\** | 101-Layer | 72.6 | 68.9 | 96.9 | 40.9 | 68.1 | 79.9 | 89.3 |
| *VGG-19\** | 19-Layer | 92.6 | 92.6 | 91.2 | 94.1 | 95.4 | 93.2 | 92.0 |
| *Xception\*[3]* | 36-Layer | 70.2 | 69.5 | 68.9 | 70.2 | N/A | N/A | N/A |

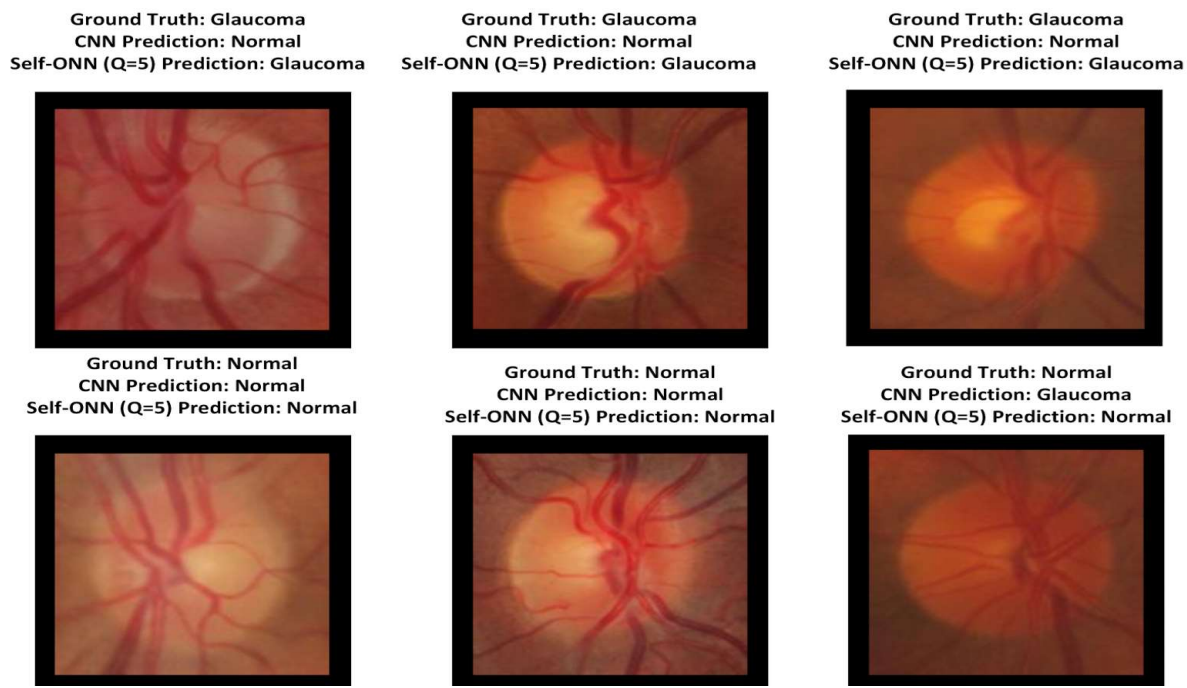

F1-score in an unfair comparison with Self-ONNs. Finally, for the ACRIMA dataset, Self-ONN with Q=5 again achieves the highest F1-score of 94%, which outperforms the equivalent CNN with a large 10% gap. For all datasets, VGG-19 achieved the second-best F1 score while the ResNet-101 and equivalent CNN shared the worst F1 performances. As for visual evaluation, a sample set of classification results on the RIM dataset are shown in Figure 4 with three pairs of normal (healthy) and glaucoma images.

A high visual similarity can be observed between the two images in each column with different classes. Such an inter-class similarity may be the main reason for the classification errors of CNN and the significant performance gap reported on each dataset. Obviously, over such a challenging pattern recognition task, conventional CNNs with the aforementioned limitations cannot perform well unless deep, complex, and pre-trained networks are used. Self-ONNs can, on the other hand, outperform even deep CNNs thanks to generative neurons.

D. Computational Complexity Analysis

To analyze computational complexity, we compute the total number of multiply-accumulate operations (MACs) and the total number of parameters (PARs) for each network configuration. The detailed formulations of the PARs and MACs calculations for Self-ONNs can be found in [23][24]. All the experiments were carried out on a 5.0 GHz Intel Core i7 with 8 GB of RAM and NVIDIA GeForce RTX 2060 graphic card. For implementation of the Self-ONNs, Python with Pytorch library is used. Both the training and testing phases of the classifier were processed using the GPU. Along with the average time complexity, we provide the overall PARs and MACs for all networks in Table V.

**TABLE V: Computational complexity levels of the networks.**

| Network | Conf. | PARs (M) | MACs (M) | Time (mSec) |
|---|---|---|---|---|
| CNN | (32+16+8) + (16+2) | 0.054 | 180.49 | 0.031 |
| Self-ONN (Q=3) | (32+16+8) + (16+2) | 0.162 | 540.57 | 0.064 |
| Self-ONN (Q=5) | (32+16+8) + (16+2) | 0.271 | 900.66 | 0.078 |
| Self-ONN (Q=7) | (32+16+8) + (16+2) | 0.379 | 1260.74 | 0.107 |
| Self-ONN (Q=9) | (32+16+8) + (16+2) | 0.488 | 1620.82 | 0.14 |
| ResNet-101* | 101-Layer | 44.55 | 2561.46 | 0.71 |
| VGG-19* | 19 Layer | 143.67 | 6622.24 | 0.99 |

In the ACRIMA dataset, the best performing Self-ONN (Q=5) requires 64.8% less MACs as compared to ResNet and 86.4% less MACs than VGG-19. This trend is consistent for other datasets too, as can be seen from Table II and Table III.

For the single-CPU implementation on an ordinary computer, the total time for the classification (FP) of a normalized input image is about 0.07 msec for a Self-ONN classifier with Q=5. Such a computation speed naturally allows real-time operation even over mobile devices with low-power CPUs.

## IV. CONCLUSIONS

In this study, Self-ONNs are proposed to detect glaucoma disease from the acquired fundus images as an alternative to the commonly applied deep CNNs with high computational complexity and special hardware requirements. The proposed classifier achieved a significant performance gap of 8-12% F1 score over equivalent CNN and even deep CNN models for the three benchmark glaucoma datasets. Deep CNNs with a large number of neurons and higher depths were outperformed by the proposed Self-ONNs despite the fact that they were pre-trained and tuned for the classification problem at hand via transfer learning. Self-ONNs achieved the state-of-the-art performance levels in glaucoma detection with a reduced complexity compared to deep CNN models, and can hence be integrated into a decision support system for real-time glaucoma detection. Future work will focus on further improving Self-ONN's classification performance by properly combining a segmentation network. The optimized PyTorch implementation of Self-ONNs is publicly shared in [41].

ACKNOWLEDGMENT

The Authors are grateful to Prof. Nilgun Yildirim and her team for providing us the labeled version of ESOGU (Eskisehir Osmangazi University) glaucoma fundus images dataset.ACKNOWLEDGMENT

The Authors are grateful to Prof. Nilgun Yildirim and her team for providing us the labeled version of ESOGU (Eskisehir Osmangazi University) glaucoma fundus images dataset.

## SUPPLEMENTARY MATERIAL

### A. Self-ONNs

In this section, we will briefly introduce the concept of ONNs and summarize the theory and forward-propagation formulations on Self-ONNs whilst the detailed derivations and Back-Propagation (BP) formulations were left to [24] and [25].

In a convolutional neuron, from each previous layer output, $y_{l-1}^j$, the input map of the $k^{th}$ convolutional neuron in layer l, $x_l^k$ is calculated as:

$$x_l^k(i,j) = \sum_{j=1}^{S_{l-1}} \sum_{u=0}^{m-1} \sum_{v=0}^{n-1} w_l^k(u,v) y_{l-1}^j(i-u, j-v) \quad (6)$$

where $y_{l-1}^j \in \mathbb{R}^{M \times N}$, $S_{l-1}$ is the number of neurons at layer l-1 and the weight kernel $w_l^k \in \mathbb{R}^{m \times n}$. For the sake of brevity, unit stride and dilation are assumed, and the input is padded with zeros before the convolution operation to preserve the spatial dimensions. An alternate formulation of the operation of (6) is now presented. Firstly, y is reshuffled such that values inside each m × n sliding block of $y_{l-1}$ are vectorized and concatenated as rows to form a matrix $Y_{l-1} \in \mathbb{R}^{\widehat{M} \times \widehat{N}}$ where $\widehat{M} = MN$ and $\widehat{N} = mn$. This operation is commonly referred to as "im2col" and is critical in conventional GEMM-based convolution implementations [23][40]. Secondly, we construct a matrix $W_l^k \in \mathbb{R}^{\widehat{M} \times \widehat{N}}$ whose rows are repeated copies of $\vec{w_l^k} = \text{vec}(W_l^k) \in \mathbb{R}^{mn}$, where $\text{vec}(\cdot)$ is the vectorization operator. Each element of $W_l^k$ is given by the following equation:

$$W_l^k(i,j) = \vec{w_l^k}(i) \quad (7)$$

The convolution operation in (6) can then be represented as follows:

$$x_l^k = \text{vec}_{M \times N}^{-1} \left( \sum_j (Y_{l-1} \otimes W_l^k) \right) \quad (8)$$

where $\otimes$ represents the Hadamard product, $\sum_j$ is the summation across $j^{th}$ dimension. In (8), $\text{vec}_{M \times N}^{-1}$ is the inverse vectorization operation that reshapes back to M × N. The formulation given in (8) can now be generically reformulated to represent the forward-propagation through an operational neuron:

$$x_l^k = \text{vec}_{M \times N}^{-1} \left( \phi_l^k ( \psi_l^k(Y_{l-1}, W_l^k) ) \right) \quad (9)$$

where $\psi(\cdot): \mathbb{R}^{M \times N} \to \mathbb{R}^{M \times N}$ and $\phi(\cdot): \mathbb{R}^{\widehat{M} \times \widehat{N}} \to \mathbb{R}^{\widehat{M}}$ are termed as nodal and pool functions [21]-[23], respectively. Finally, after applying the activation function $f_l^k$, we get the output of the neuron:

$$y_l^k = f_l^k \left( \text{vec}_{M \times N}^{-1} \left( \phi_l^k \left( \psi_l^k(Y_{l-1}, W_l^k) \right) \right) \right) \quad (10)$$

Given an operator set; a triplet of $(\psi_l^k, \phi_l^k, f_l^k)$, an operational neuron implements the formulation given in (10). It can be noticed here that the convolutional neuron is a special case of an operational neuron with nodal function $\psi(\alpha, \beta) = \alpha * \beta$ and pooling function $\phi(\cdot) = \sum_i$.

In a heterogenous ONN configuration, every neuron has uniquely assigned ψ and P operators. Owing to this, an ONN network enjoys the flexibility of incorporating any non-linear transformation, which is suitable for the given learning problem at hand. However, hand-crafting a suitable library of possible operators and searching for an optimal one for each neuron in a network introduces a significant overhead, which rises exponentially with increasing network complexity. Moreover, it is also possible that the right operator for the given learning problem cannot be expressed in terms of well-known functions. Self-ONN has the potential to achieve a superior operational diversity and flexibility that permits the formation of any nodal operator function without the need of any operator set library or any prior search process to find the optimal nodal operator.

To formulate a nodal transformation, ψ which does not require a pre-selection and manual assignment of operators, we use the Taylor-series based function approximation near the origin (a = 0), as,

$$\psi(x) = \sum_{n=0}^{\infty} \frac{\psi^{(n)}(0)}{n!} x^n \quad (11)$$

The $Q^{th}$ order truncated approximation, formally known as the Taylor polynomial, takes the form of the following finite summation:

$$\psi(x)^{(Q)} = \sum_{n=0}^{Q} \frac{\psi^{(n)}(0)}{n!} x^n \qquad (12)$$

The above formulation can approximate any function $\psi(x)$ sufficiently well near 0. When the activation function bounds the neuron's input feature maps in the vicinity of 0 (e.g., tanh) the formulation of (2) can be exploited to form a composite nodal operator where the power coefficients, $\frac{\psi^{(n)}(0)}{n!}$ can be the learned parameters of the network during training. It was shown in [25] that the nodal operator of the kth generative neuron in the lth layer can take the following general form:

$$\psi_l^k(Y_{l-1}, W_l^k, Q) = \sum_{q=1}^{Q} Y_{l-1}^q \otimes W_l^{k\,(q)} \qquad (13)$$

where $W_l^k \in \mathbb{R}^{\widehat{M} \times \widehat{N} \times Q}$ is the three-dimensional weight matrix and $W_l^{k(q)} \in \mathbb{R}^{\widehat{M} \times \widehat{N}}$ is the $q^{th}$ slice of $W_l^k$. The $0^{th}$ order term a, the DC bias, is ignored as its additive effect can be compensated by the learnable bias parameter of the neuron. Back-propagation (BP) through this nodal operator is now trivial to accomplish. Equations (14) and (6) provide the derivatives with respect to the input $Y_{l-1}$ and the $q^{th}$ slice weights, $W_l^{k\,(q)}$, respectively:

$$\frac{d\psi_l^k}{dY_{l-1}} = \sum_{q=1}^{Q} q Y_{l-1}^{q-1} \otimes W_l^{k\,(q)} \qquad (14)$$

$$\frac{d\psi_l^k}{dW_l^{k\,(q)}} = Y_{l-1}^q \qquad (6)$$

The detailed formulations of the BP training in raw-vectorized form can be referred to [25].[23]